\documentclass[
 reprint,
 amsmath,amssymb,
 aps,
 pra,
]{revtex4-2}

\usepackage{graphicx}
\usepackage{dcolumn}
\usepackage{bm}

\usepackage{physics}
\usepackage{xcolor}
\usepackage{xparse}

\usepackage{mathtools}

\newcommand{\tab}{\ \ \ \ }
\newcommand{\half}{\frac{1}{2}}
\newcommand{\rt}{\frac{1}{\sqrt{2}}}

\renewcommand{\a}{\hat{a}}

\newcommand{\id}{\openone}

\begin{document}

\preprint{APS/123-QED}

\title{Power-optimized amplitude modulation for robust trapped-ion entangling gates: \\a study of gate-timing errors}

\author{Luke Ellert-Beck}
 \email{lellertbeck@uri.edu}
\author{Wenchao Ge}
\email{wenchao.ge@uri.edu}
\affiliation{
 Department of Physics, University of Rhode Island, Kingston, RI 02881 \\
 Department of Physics, Southern Illinois University, Carbondale, IL 62901 
}

\date{\today}

\begin{abstract}
Trapped-ion systems are a promising route toward the realization of both near-term and universal quantum computers.
However, one of the pressing challenges is improving the fidelity of two-qubit entangling gates. These operations are often implemented by addressing individual ions with laser pulses using the M\o lmer-S\o rensen (MS) protocol. 
Amplitude modulation (AM) is a well-studied extension of this protocol, where the amplitude of the laser pulses is controlled as a function of time.
We present an analytical study of AM, using a Fourier series expansion to maintain the generality of the laser amplitude's functional form. 
We then apply this general AM method to gate-timing errors by imposing conditions on these Fourier coefficients, producing trade-offs between the laser power and fidelity at a fixed gate time.
The conditions derived here are linear and can be used, in principle, to achieve arbitrarily high orders of insensitivity to gate-timing errors.
Numerical optimization is then employed to identify the minimum-power pulse satisfying these constraints.
Our central result is that the leading order dependence on gate timing errors is improved from $\order{\Delta t^2}$ to $\order{\Delta t^6}$ with the addition of one linear constraint on the Fourier coefficients and to $\order{\Delta t^{10}}$ with two linear constraints without a significant increase in the average laser power. The increase approaches zero as more Fourier coefficients are included. 
In further studies, this protocol can be applied to other error sources and used in conjunction with other error-mitigation techniques to improve two-qubit gates.
\end{abstract}

\maketitle

\section{INTRODUCTION}

Trapped-ion systems are promising candidates for near-term and universal quantum computers. They have long coherence times, up to 0.2 s for optical qubits and up to 600 s for hyperfine qubits~\cite{bruzewicz-ti-review}, and support high-fidelity operations, with single-qubit gate fidelities exceeding $0.9999$~\cite{HartyPRL2014} and two-qubit gate fidelities exceeding $0.999$~\cite{BallancePRL2016, GTL+16, ClarkPRL2021}. While three nines (.999) in two-qubit gate fidelities are sufficient for fault-tolerant quantum computation (see, for example, the .9925 threshold~\cite{RH07}), any additional progress will significantly reduce the overhead required for quantum error correction. Therefore, improving two-qubit gate fidelities will increase the range of computations available on quantum information processors, particularly when the system scales up to more qubits.

The Molmer-Sorensen (MS) gate~\cite{ms-pra,ms-prl} is one of the standard two-qubit gates, with fidelities exceeding .999 in experimental implementations~\cite{GTL+16,bruzewicz-ti-review}. Although these figures place the MS gate within the fault-tolerance threshold, the fidelity suffers greatly from experimental imperfections, such as motional mode frequency errors, laser (or qubit) frequency fluctuations, and gate-timing errors, especially as the system scales.

Modifications of the MS protocol, such as pulse shaping, have been proposed to reduce its sensitivity to these fluctuations.
Each of the three pulse parameters—amplitude~\cite{HC+12,CD+14,ZH+19,efficient,power-optimal,JH+23,DZ+22,RC+22-pub24}, frequency~\cite{LL+18,KL+21,efficient,power-optimal,JH+23,KW+23}, and phase~\cite{GB15,ME+20}—can be controlled as time-dependent functions to enhance the protocol. Additionally, pulses can be constructed from multiple frequency tones~\cite{robust, WW+18,SS+20,LM20}. These degrees of freedom allow pulses to be tailored toward certain outcomes, such as robustness to errors. 

Amplitude modulation (AM) has been demonstrated experimentally in different forms in both laser-based and laser-free setups. AM can be implemented with discrete segments~\cite{HC+12,CD+14} or by using smooth envelope functions~\cite{ZH+19,RC+22-pub24,DZ+22}. Common pulse envelopes for AM are the $\sin^2$ pulse~\cite{ZH+19} and the truncated Gaussian pulse~\cite{RC+22-pub24}. These pulses feature a `soft start'~\cite{ZH+19}, which results in a smaller displacement near the gate time. Thus, these gates are more robust to errors caused by residual spin-motion entanglement. Another strategy is to use a parameterized pulse shape to bring out desired features (such as robustness to experimental parameters or smoothness). In a further study of~\cite{ZH+19}, a general pulse shape composed of triangular basis functions was used~\cite{DZ+22}. 

In this work, we employ the general method of pulse amplitude modulation using a Fourier basis to describe our family of functions.
The general method is then applied to gate-timing errors.
The robustness of the gate against timing errors is guaranteed by a sequence of linear equations, whose explicit forms are derived in terms of the Fourier coefficients.
In particular, the leading-order sensitivity is asymptotically improved by $\Delta t^4$ (for a gate time error $\Delta t$) for each linear constraint used. 
The power-optimized pulse is then selected by reduction to a generalized eigenvalue problem, which is solved numerically.
This benefit is observed with minimal additional laser power, as compared to the original MS protocol, if sufficiently many frequencies are used in the AM.

A similar approach, using Fourier pulse shaping and laser power optimization, was developed in~\cite{power-optimal,efficient}. They perform amplitude and frequency modulation simultaneously and represent the total pulse (the amplitude envelope and a frequency modulating term) with a Fourier sine series. The key features of their scheme are highlighted here. Primarily, they use a more complicated model, where all motional modes of a chain of $N$ ions are considered, and fast-oscillating terms are kept. This allows their work to achieve a practical improvement to the fidelity of current devices. Furthermore, their method produces a provably power-optimal pulse using matrix operations. Finally, while they provide the form of the linear constraints used to achieve robustness to timing errors in the supplemental materials of Ref.~\cite{power-optimal}, they mainly use the method to analyze the closure of multiple phase space loops in the presence of motional mode frequency errors.

In our scheme, we consider two ions interacting with only the center-of-mass mode. 
This simple model allows us to derive analytical expressions for the phase space trajectories of MS gates for AM generally, with pulses characterized by Fourier coefficients. 
Power optimization is required to minimize the cost, and a similar optimization method based on matrix operations is employed in our protocol.
To include the robustness to gate-timing errors, we derive a set of linear equations in terms of the Fourier coefficients which, when satisfied, provides this benefit. 
Applying a general method of AM to a simple system is also useful for investigating whether such improvements to the MS protocol can be attained with vanishing cost. We find evidence that this can be done as the number of Fourier coefficients increases.
In addition, we demonstrate these goals with only tens of basis functions, which is about $2$ orders of magnitude fewer than the five-qubit example in Ref.~\cite{power-optimal}.

This work is organized as follows. Section~\ref{sec:ms} introduces the MS gate in more details, using the notation and assumptions considered in~\cite{ms-pra}. 
In Section~\ref{sec:methods}, the method is described in detail. We define the Fourier series structure of our AM (\ref{sec:GAM}) and derive sensitivity conditions for gate-timing errors (\ref{sec:ExpFG}-C). Then, we perform optimization of the free parameters to reduce the power cost (\ref{sec:PO}).
Section~\ref{sec:results} contains our numerical results, including plots of the power-optimized pulse envelopes, gate trajectories, and population evolution. Section~\ref{sec:conclusion} contains our discussion and outlook.

\section{The M\o lmer-S\o rensen Gate} \label{sec:ms}
\subsection{Introduction} \label{sec:ms-intro}
One well-established protocol for implementing entangling gates on trapped-ion platforms is the M\o lmer-S\o rensen (MS) gate \cite{ms-pra}, which uses the ions' motional modes to mediate an effective spin-spin interaction between the internal electronic states, i.e., the qubits. 
The MS gate, or MS protocol, is realized by addressing the ions with bichromatic laser light, with beams symmetrically detuned from the qubit transition frequency.

In this work, we consider the situation of two ions in a linear trap and assume that both ions have the same Rabi frequency $\Omega$ and Lamb-Dicke parameter $\eta$. We also assume that the trap is cooled to the Lamb-Dicke regime, meaning that motional excitations greater than one quanta are suppressed. 
In addition, we take the symmetric laser detuning $\delta$ to be close to the center-of-mass mode frequency $\nu$ ($\nu-\delta\ll\delta$), which allows us to consider only the center-of-mass mode in the interaction. Assuming this and that the laser intensity is not too large ($\Omega\ll\delta$), the carrier term (recoil-free atomic transitions) and fast-oscillating terms (those that oscillate at frequencies $\delta$ or $\nu+\delta$) are neglected. 

The starting point for this work is the MS Hamiltonian in the interaction picture. Setting $\hbar=1$, working in the Lamb-Dicke regime, and neglecting carrier transitions and the fast-rotating terms, the interaction picture Hamiltonian is~\cite{ms-pra}
\begin{align}
    H_\text{int} &= f(t) J_y x + g(t) J_y p,
\end{align}
where $J_y=( \sigma_y^1+ \sigma_y^2)/2$ is the collective spin operator with Pauli operator $\sigma_y$, $x$ and $p$ are the position and momentum operators, respectively, of the center-of-mass mode, and the functions $f$ and $g$ are defined as
\begin{align}
    f(t) &= -\sqrt{2}\eta\Omega\cos(\xi_0t), \\
    g(t) &= -\sqrt{2}\eta\Omega\sin(\xi_0t).
\end{align}
Here $\xi_0\equiv\nu-\delta$ is the laser detuning from the red and blue sidebands of the center-of-mass motion.

This Hamiltonian gives us an exact propagator~\cite{ms-pra}
\begin{align}
    U(t) &= e^{-i A(t) J_y^2} e^{-i F(t) J_y x} e^{-i G(t) J_y p}, \label{ms-prop}
\end{align}
which introduces the functions $F,G,$ and $A$, defined by
\begin{align}
    F(t) &= \int_0^t\dd{t'}f(t'), \label{F-def}\\
    G(t) &= \int_0^t\dd{t'}g(t'), \label{G-def}\\
    A(t) &= -\int_0^t\dd{t'}F(t')g(t'). \label{A-def}
\end{align}
At the gate time $T= 2\pi/\xi_0$ under ideal circumstances, we get
\begin{align}
    F(T) = G(T) &= 0, \label{closure} \\
    A(T) &= \pi/2. \label{phase}
\end{align}
Equation \eqref{closure} is the condition when the spins and the center-of-mass motion decouple; equation~\eqref{phase} is the condition for preparing the maximally entangled two-qubit state. These conditions, applied to~\eqref{ms-prop}, yield the propagator
\begin{align*}
    U(T) &= e^{-i\frac{\pi}{2} J_y^2},
\end{align*}
which generates the YY-interaction between the two ions. Let $\ket{gg}$ be the state where both two-level atoms are in the ground state and $\ket{ee}$ be the state where both atoms are in the excited state. Assuming the initial state to be $\ket{gg}$, the protocol yields the final state
\begin{align}
    \ket{\psi_\text{ideal}} &= U(T)\ket{gg} = \rt\left(\ket{gg}+i\ket{ee}\right), \label{psi-ideal}
\end{align}
under ideal conditions. This logical gate is a sufficient entangling gate for universal computing in combination with arbitrary single-qubit gates~\cite{elem-gates}.

\begin{figure}
    \includegraphics[width=0.35\textwidth]{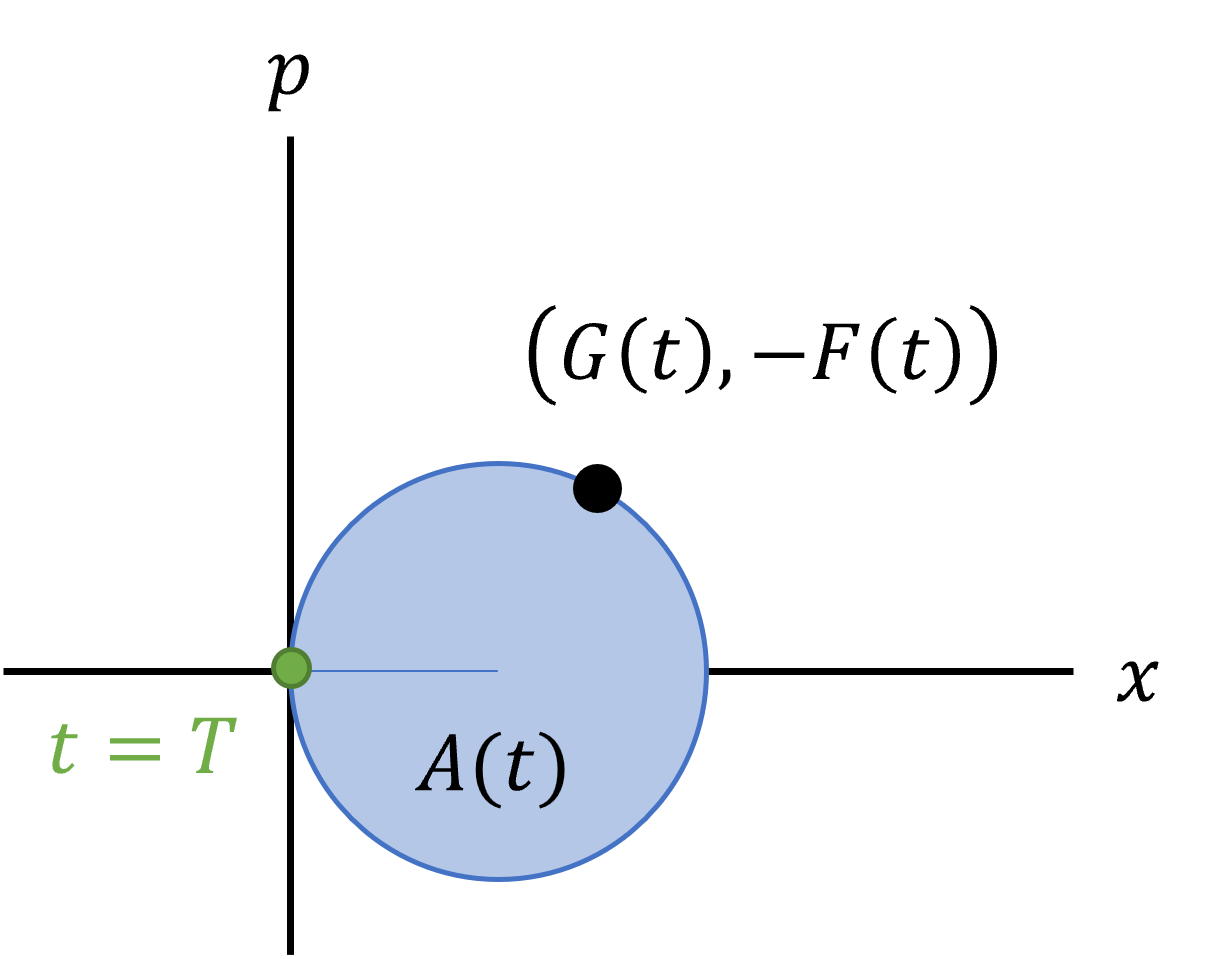}
    \caption{Phase space picture of the MS gate with no error. The variables $x$ and $p$ are the position and momentum observables of the center-of-mass oscillation of the two-ion system. $F(t)$ and $G(t)$ determine the spin-motion coupling as a function of time $t$ and therefore determine the phase space trajectory. As a result, the system acquires a geometric phase $A(t)$, particularly at the gate time $T$ when the trajectory closes at the origin and the spin variables decouple from the motion. }
    \label{fig:ps-compact}
\end{figure}

We note that the propagator in Eq.~\eqref{ms-prop} acts as a displacement operator in the phase space of the motional mode, so the phase space picture will be used to visualize the gate trajectories, as shown in Figure \ref{fig:ps-compact}. In this geometric interpretation, the $x$- and $p$-coordinates of the phase space trajectory are given directly by the functions $G(t)$ and $-F(t)$, respectively. The geometric phase $A(t)$ is the area enclosed by the trajectory. The MS protocol generates a circular trajectory in phase space which begins at the origin and closes at the origin exactly one gate time $T$ later. This closure at the origin is critical to creating a maximally entangled state in the electronic states of the ions, because a displacement from the origin in phase space represents a coupling between the motional modes and the spin variables. This remaining coupling prevents the spin states from becoming maximally entangled, which reduces the gate fidelity.

\subsection{Gate Fidelity} \label{sec:ms-fid}
Our aim is to modify the MS protocol so that it produces an entangled state at a predetermined level of success in the presence of noise and imperfections. We quantify the success of the protocol by the gate fidelity---the projection of the prepared state onto the ideal state---defined by
$$
    \expval{\hat\rho(T)}{\psi_\text{ideal}},
$$
where $\hat\rho(T)$ is the density matrix of the state prepared after one gate time $T$, and $\ket{\psi_\text{ideal}}$ is the desired state. 

Reference \cite{robust} gives the following analytical expression for the MS gate fidelity, 
\begin{align}
    \mathcal{F}_{MS} &= \frac{3+e^{-4(\bar{n}+\half)\frac{F^2+G^2}{2}}}{8} \nonumber \\ &\tab + \frac{e^{-(\bar{n}+\half)\frac{F^2+G^2}{2}}\sin(A+\frac{FG}{2})}{2}, \label{fidelity}
\end{align}
where $\bar{n}$ is the average phonon number of the initial motional state, which is directly related to the initial temperature of the system, $F$ and $G$ are the phase space coordinates~(\ref{F-def}, \ref{G-def}), and $A$ is the geometric phase~\eqref{A-def}. Note that the quantity $\bar n$ is not a parameter in our study, as it is not related to the pulse shape. Experimentally, this quantity can be made very small by cooling the ions near the ground state motional mode~\cite{bruzewicz-ti-review,cooling-nbar}.

In the ideal case, when the gate time is exact and there are no fluctuations in the other parameters, $F=0$, $G=0$, and $A=\pi/2$. Hence, $\mathcal{F}_{MS}=1$, and the state $\ket{\psi_\text{ideal}}$ is prepared exactly. Other parameters, such as motional mode frequency, laser frequency, laser intensity, and gate time, affect $F, G,$ and $A$. Therefore, they also influence the fidelity. In this work, we focus on reducing errors due to imperfections in the gate time as an example of the power-optimized amplitude modulation protocol.

\subsection{Expansion of the Gate Fidelity} \label{sec:ms-exp}
Though the exact expression for fidelity is known in Eq.~\eqref{fidelity}, an expansion is used to analyze the sensitivity of the relevant quantities to gate timing errors.
Consider a trajectory $(G,-F)$ which closes at the origin and periodic functions $f(t),g(t)$. A variation in the gate time $t=T+\Delta t$ gives rise to the following variations in $F,G,$ and $A$, which follow from the linearity of the integral expressions,
\begin{align}
    \Delta F \equiv F(T+\Delta t) - F(T) &= F(\Delta t), \label{GplusDel} \\
    \Delta G \equiv G(T+\Delta t) - G(T) &= G(\Delta t), \label{FplusDel} \\
    \Delta A \equiv A(T+\Delta t) - A(T) &= A(\Delta t) - \pi/2. \label{AplusDel}
\end{align}

If we consider only small variations in the gate time, i.e. $\Delta t\ll T$, then the variations in phase space coordinates and geometric phase should also be small, $\Delta F\ll 1$, $\Delta G\ll 1$, and $\Delta A - \pi/2 \ll 1$. It follows that the infidelity is, to leading order (see Appendix~\ref{app:expansion-details}),
\begin{align}
    1-\mathcal{F}_{MS}\approx\left(\bar{n}+\half\right)\frac{F^2+G^2}{2},
    \label{eq:infidelity}
\end{align}
as long as $\bar n$ is not too large.

In the original MS protocol \cite{ms-pra}, recall $F(t) \propto \sin(\xi_0 t)$, and $G(t) \propto 1-\cos(\xi_0 t)$. Hence, for gate time fluctuations,
\begin{align*}
    \Delta F &\propto \sin(\xi_0\Delta t) = \order{\Delta t},
\end{align*}
while
\begin{align*}
    \Delta G &\propto 1-\cos(\xi_0\Delta t) = \order{\Delta t^2},
\end{align*}
which means that $F^2+G^2 = \order{\Delta t^2}$ and hence the leading order term of the fidelity of the traditional MS gate is $\order{\Delta t^2}$. Throughout the rest of this article, a protocol capable of improving the leading contribution to arbitrary order is developed analytically. The next two sections contain the methods used and the numerical results for examples of $\order{\Delta t^6}$ and $\order{\Delta t^{10}}$ sensitivity.

\section{METHODS} \label{sec:methods}
In this work, we apply amplitude modulation to the two-qubit MS gate, constrain the asymptotic scaling of the MS gate fidelity with respect to gate-timing errors, then perform numerical optimization to select the minimal-power pulse which satisfies these constraints.

The AM is treated analytically by considering a Fourier series expansion of the pulse envelope function. 
This accomplishes a description of all continuous pulses, which are characterized by the Fourier coefficients.
An analytical (parametric) expression for the phase space trajectory of the gate is derived in terms of these coefficients, which characterizes the effect of the gate in terms of the Fourier coefficients.

As in~\cite{robust}, we use a series expansion of the MS gate fidelity with respect to a small gate time deviation $\Delta t$ to derive constraints for the sensitivity. However, the leading order dependence of the fidelity to variations in $F$ and $G$ were determined in the previous section, so all that remains is to show the dependence of $F$ and $G$ on gate time deviations. This is achieved using a Taylor series expansion, in which the $\order{\Delta t^n}$ contribution of $F$ (and $G$) is determined by its $n$th derivative. The leading order term can thus be chosen to arbitrary order: setting the first $k$ derivatives of $F$ and $G$ to zero results in a gate fidelity with leading order $\order{\Delta t^{2k+2}}$.
These constraints are derived analytically and reduced to an equivalent system of linear equations on the Fourier coefficients used to express the pulse envelope function. Moreover, we find that it is optimal to set the coefficients of the odd basis functions to zero, meaning that half of the linear constraints (LCs) would be automatically satisfied. Hence, a system of $l$ LCs are used to improve the leading order gate fidelity to $\order{\Delta t^{4l+2}}$---an improvement of four orders for each LC added.

Considering the power cost, we find that the average laser power and geometric phase are both quadratic in terms of the Fourier coefficients. Therefore, a pulse with improved fidelity and minimized laser power can be found by solving a constrained quadratic optimization problem. We solve the optimization problem by reducing it via the linear constraints, then changing coordinates so that the second quadratic constraint is normalized. This problem is reducible to the eigenvalue problem, so the relevant matrix is constructed analytically (index-wise) then diagonalized numerically. From these eigenvectors, the $a_i$ Fourier coefficients are recovered and the power-optimized pulse is constructed.

\subsection{General Amplitude Modulation} \label{sec:GAM}
The procedure begins with a general time-dependent amplitude envelope function $\Omega(t)$ written in the Fourier basis. This can be made equivalent to the simultaneous amplitude modulation and frequency modulation discussed in Ref.~\cite{power-optimal}, where a fast-oscillating pulse function is defined as $\mathcal{G}(t)=\Omega(t)\sin[\psi(t)]$ to incorporate both the envelope function of the pulse $\Omega(t)$ and the phase term $\psi(t)$. We argue that by expanding only the amplitude $\Omega(t)$ in the complete basis, one can generate the same pulse function $\mathcal{G}(t)$ in the fast-rotating limit as long as the basis expansion converges to $\mathcal{G}(t)/\sin[\psi(t)]$. 

We express the amplitude envelope function $\Omega(t)$ as a Fourier series with coefficients $a_0, a_n, b_n$, so that
\begin{align}
    \Omega(t) &= \frac{a_0}{2} + \sum_{n=1}^N a_n\cos(n\xi_0t)+b_n\sin(n\xi_0t).\label{gen-fs}
\end{align}
Note that we truncate the series to $N$th order in this expression, yet the analysis holds for arbitrarily large $N$ in principle. The coefficients $a_n, b_n$ give us a large parameter space to work with, which will be used to characterize AM pulses generally, impose robustness constraints, and optimize the laser power.
The derivation of the MS propagator \eqref{ms-prop} is unchanged, however, our expressions $f(t)$ and $g(t)$ gain the time-dependence
\begin{align}
    f(t) &= -\sqrt{2}\eta\Omega(t)\cos(\xi_0t), \label{f-td} \\
    g(t) &= -\sqrt{2}\eta\Omega(t)\sin(\xi_0t), \label{g-td}
\end{align}
The integration to find $F(t)$ and $G(t)$ can still be done analytically with the choice of the Fourier basis. So, the exact gate trajectory for AM in the Fourier basis is given by
\begin{widetext}
\begin{align}
    F(t) &= -\frac{\eta}{\sqrt{2}\xi_0} \Bigg[ a_0\sin(\xi_0t) + a_1\left(\frac{\sin(2\xi_0t)}{2}+\xi_0t\right)  + 
    \sum_{n=2}^N a_n\left(\frac{\sin((n+1)\xi_0t)}{n+1} + \frac{\sin((n-1)\xi_0t)}{n-1}\right) \nonumber\\&\tab\tab\tab\tab+ 
    b_1\left(\frac{1-\cos(2\xi_0t)}{2}\right) + \sum_{n=2}^N b_n\left(\frac{1-\cos((n+1)\xi_0t)}{n+1}+\frac{1-\cos((n-1)\xi_0t)}{n-1}\right)\Bigg], \label{F-major}
\end{align}
and
\begin{align}
    G(t) &= -\frac{\eta}{\sqrt{2}\xi_0} \Bigg[ a_0(1-\cos(\xi_0t)) + a_1\left(\frac{1-\cos(2\xi_0t)}{2}\right) 
    +
    \sum_{n=2}^Na_n\left(\frac{1-\cos((n+1)\xi_0t)}{n+1}-\frac{1-\cos((n-1)\xi_0t)}{n-1}\right) 
    \nonumber\\ &\tab\tab\tab\tab+ b_1\left(\xi_0t-\frac{\sin(2\xi_0t)}{2}\right) + \sum_{n=2}^Nb_n\left(\frac{\sin((n-1)\xi_0t)}{n-1}-\frac{\sin((n+1)\xi_0t)}{n+1}\right)\Bigg]. \label{G-major}
\end{align}
\end{widetext}
Note that $a_1$ and $b_1$ are each coefficients of a term linear in $t$ [in~\eqref{F-major} and~\eqref{G-major}, respectively]. Therefore, $F(T)=G(T)=0$ implies $a_1=b_1=0$. These exact expressions for the gate trajectories are used in Section~\ref{sec:LCs} to exclude $n=1$ from the optimization and in Section~\ref{sec:results-am-vs-ms} to speed up numerical calculations.

\subsection[Expansion of F and G]{Expansion of $F$ and $G$} \label{sec:ExpFG}
The infidelity is proportional to $F^2+G^2$ to the leading order~\eqref{eq:infidelity}, thus the sensitivity of the fidelity to experimental parameters is determined by constraints on the derivatives of $F$ and $G$. In particular, our strategy is to declare a leading order $k+1$ such that $\Delta F = \order{\Delta t^{k+1}}$ and $\Delta G = \order {\Delta t^{k+1}}$, which guarantees sensitivity to errors of $1-\mathcal{F}_{MS}=\order{\Delta t^{2k+2}}$.

The general forms~\eqref{F-major} and~\eqref{G-major} could be used to calculate the derivatives, but the Taylor expansions of $F$ and $G$ themselves are convenient to work with. Due to our choice of the Fourier basis for the expansion of $\Omega(t)$, the derivatives of $F$ and $G$ are readily calculated analytically in terms of the Fourier coefficients $a_n$ and $b_n$. For this reason, we'll derive the conditions for the function $\Omega(t)$ from these Taylor expansions directly. Recall the definitions of $F$ and $G$ (equations \eqref{F-def} and \eqref{G-def}, respectively), which directly yield $F'(t)=f(t)$ and $G'(t)=g(t)$ by the fundamental theorem of calculus. Thus we just need to examine the derivatives of $f(t)$ and $g(t)$, defined in equations \eqref{f-td} and \eqref{g-td}, respectively.

We make use of the general Leibniz rule, which states
\begin{align*}
    (fg)^{(n)} &= \sum_{k=0}^n \binom{n}{k} f^{(n-k)}g^{(k)}
\end{align*}
for arbitrary functions $f$ and $g$. Therefore,
\begin{align}
    f^{(i)}(t) &= -\sqrt{2}\eta \sum_{l=0}^i \binom{i}{l} \Omega^{(i-l)}(t)\pdv[l]{t}\big(\cos(\xi_0 t)\big) \nonumber \\ 
    &= -\sqrt{2}\eta \sum_{l=0}^i (\xi_0)^l \binom{i}{l} \Omega^{(i-l)}(t)\cos(\xi_0 t + l\pi/2),\label{f-derivs}
\end{align}
and 
\begin{align}
    g^{(i)}(t) &= -\sqrt{2}\eta \sum_{l=0}^i \binom{i}{l} \Omega^{(i-l)}(t)\pdv[l]{t}\big(\sin(\xi_0 t)\big) \nonumber \\
    &= -\sqrt{2}\eta  \\ &\tab\times \sum_{l=0}^i (\xi_0)^l \binom{i}{l} \Omega^{(i-l)}(t)\cos(\xi_0 t + (l-1)\pi/2). \label{g-derivs}
\end{align}
These expressions depend on the derivatives of $\Omega(t)$, which are described in Appendix~\ref{app:Omega-derivatives}. After simplifying the expression at the gate time $t=T$, the $i$th order derivatives of $F$ and $G$ are
\begin{widetext}
\begin{align}
    F^{(i)}(T) &= \begin{dcases}
        (-1)^{\frac{i+1}{2}}\sqrt{2}\eta\xi_0^{i-1}\left[\frac{a_0}{2}+\sum_{n=1}^N a_n \sum_{l=0}^\frac{i-1}{2} \binom{i-1}{2l} n^{i-1-2l}\right], & \text{odd } i, \\
        (-1)^\frac{i}{2}\sqrt{2}\eta\xi_0^{i-1} \sum_{n=1}^N b_n \sum_{l=0}^{\left(\frac{i}{2}-1\right)} \binom{i-1}{2l} n^{i-1-2l}, & \text{even } i>0,
    \end{dcases} \label{F-i}
\end{align}
and
\begin{align}
    G^{(i)}(T) &= \begin{dcases}
        (-1)^\frac{i}{2}\sqrt{2}\eta\xi_0^{i-1}\left[\frac{a_0}{2}+\sum_{n=1}^N a_n \sum_{l=0}^\frac{i-2}{2}\binom{i-1}{2l} n^{i-2-2l}\right], & \text{even } i>0, \\
        (-1)^\frac{i-1}{2}\sqrt{2}\eta\xi_0^{i-1}\sum_{n=1}^N b_n \sum_{l=0}^\frac{i-3}{2} \binom{i-1}{2l} n^{i-2-2l}, & \text{odd } i.
    \end{dcases} \label{G-i}
\end{align}
\end{widetext}

This provides a general way to write and determine the conditions for fidelity improvements in terms of the Fourier coefficients.

\subsection{Linear Constraints} \label{sec:LCs}
The leading contribution to the fidelity is $\order{\Delta t^{2k+2}}$ when the first $k$ derivatives of $F$ and $G$ are zero. Thus, the linear system 
\begin{align}
    F^{(i)}(T) &= 0, & 0\leq i\leq k, \label{f-constraint}\\
    G^{(i)}(T) &= 0, & 0\leq i\leq k, \label{g-constraint}
\end{align}
constrains the pulse such that the fidelity will be of order $2k+2$. In Appendix~\ref{app:LCs}, we show that the constraints defined by the system given in equations \eqref{F-i} and \eqref{G-i} to satisfy \eqref{f-constraint} and \eqref{g-constraint} are captured by the reduced system
\begin{align}
    \delta_{1i}\frac{a_0}{2} + \sum_{n=1}^N a_n n^{i-1} &= 0, & \text{ for odd } i>0, \label{red-odd-l} \\
    \sum_{n=1}^N b_n n^{i-1} &= 0, &  \text{ for even } i>0, \label{red-even-l}
\end{align}
where the Kronecker delta is used for compact notation. 

In matrix form, the constraints~\eqref{red-odd-l} and~\eqref{red-even-l} can be written
\begin{align*}
    \mqty( \half & 1 & 1 & 1 & \cdots & 0 & 0 & 0 & \cdots \\
    0 & 1 & 4 & 9 & \cdots & 0 & 0 & 0 & \cdots \\
    0 & 0 & 0 & 0 & \cdots & 1 & 2 & 3 & \cdots)\smqty(a_0 \\ a_1 \\ \vdots \\ a_N \\ b_1 \\ \vdots \\b_N) &= 0,
\end{align*}
for the system up to $i=3$. This matrix, denoted as $\vb C$, can be constructed for any $i$. 
Once we decide on an order $k$ to bound the error with respect to gate time, we immediately know the number of constraints in our system $i$, and hence we know the full matrix $\vb C$. 

Note that $F(T)=G(T)=0$ implies $a_1=b_1=0$. Therefore, $n=1$ is excluded throughout the remainder of the calculation. 

\subsection{Power Optimization} \label{sec:PO}
The geometric phase and the average laser power of the gate are evaluated at the gate time to produce expressions of these physical quantities in terms of the Fourier coefficients. These expressions allow us to optimize the average power used during the gate. The optimization problem consists of extremizing a quadratic form with respect to one quadratic constraint and a collection of linear constraints, the generic solution of which can be found in reference~\cite{golub1970stationary}. 

The expression for average power is obtained by integrating $\abs{\Omega(t)}^2$ over one period; the expression for geometric phase comes from evaluating equation (\ref{A-def}) at the gate time by integration. The most general forms, in terms of both $a_i$ and $b_i$, are presented in Appendix~\ref{app:b=0}, however, we find that it is beneficial to have the $b_i$ coefficients equal to zero (see Appendix~\ref{app:b=0}). In terms of only $a_n$ coefficients, the expression for average power is
\begin{align}
    P &= \frac{1}{4}a_0^2 + \half\sum_{n=2}^N a_n^2, \label{power-a}
\end{align}
and the expression for geometric phase is
\begin{align}
    A &= -\frac{1}{4}a_0^2 + \half \sum_{n=2}^N \frac{a_n^2}{n^2-1}. \label{phase-a}
\end{align}
The above equations lead to the minimum average power usage to generate the maximally-entangled qubit states, which is $P\ge \pi/2$, which agrees with the lower bound obtained in Ref.~\cite{power-optimal} (using the definition from~\cite{CD+14}). Note that the average power and geometric phase are both quadratic forms of the Fourier coefficients, giving rise to the quadratic optimization problem. The additional linear constraints are those derived in Section~\ref{sec:LCs} that improve the sensitivity of the gate fidelity. 

Consider the row vector of $a_n$ coefficients,
\begin{align}
    \vb a^T &\equiv (a_0, a_2, \dots, a_N),
\end{align}
excluding $a_1$, which we know must be zero to ensure the gate's closure in phase space. Now we can write
\begin{align}
    P = \vb a^T\vb P \vb a,
\end{align}
with \begin{align*}
    \vb P &= \mqty(\dmat[0]{\frac{1}{4},\half,\ddots,\half}).
\end{align*}
and $A$ can be written in the matrix form as
\begin{align}
    \vb a^T\vb A\vb a
\end{align}
with
\begin{align*}
    \vb A &= \mqty(\dmat[0]{-\frac{1}{4},\frac{1}{6},\ddots,\frac{1}{2(N^2-1)}}).
\end{align*}

In terms of these quantities, the optimization problem we face is the following:
\begin{align}
    \text{maximize}   &\quad \abs{\frac{\vb a^T\vb A\vb a}{\vb a^T\vb P\vb a}} \\
    \text{subject to} &\quad \vb C\vb a = \vb 0. \nonumber
\end{align}
In principle, the geometric phase $A$ should be fixed [so that the ideal state \eqref{psi-ideal} is correctly prepared], so the average laser power $P$ should be minimized. However, the constraint for average power gives a positive definite quadratic form, whereas the geometric phase gives an indefinite one. Therefore, it is necessary to \textit{maximize} the magnitude of $A$ with respect to a fixed $P$ and rescale the solutions $\vb a$ to yield the ideal state \footnote{We anticipated (and confirmed) that the $a_0$ term would be the dominant contribution to the geometric phase and thus that $A$ will be negative. Hence, in practice, the minimum eigenvalue of $A$ with respect to $P$ was found}.

We can eliminate the linear constraints by solving the system $\vb C\vb a = \vb 0$. The solution space, the null space of $\vb C$, has dimension $N-k/2$, and so each $N$-dimensional vector $\vb a$ can be described by a unique $(N-k/2)$-dimensional vector $\vb a'$~\footnote{The strategy, used in~\cite{golub1970stationary}, is to choose vectors with $N$ entries where the last $k/2$ are zeros. Then, only the $(N-k/2)\times(N-k/2)$ submatrices which act on the nonzero entries need to be considered}. When the problem is restated in terms of these vectors, we have
\begin{align}
    \text{maximize}   &\quad \abs{\frac{\vb a'^T\vb A'\vb a'}{\vb a'^T\vb P'\vb a'}},
\end{align}
where the linear constraints are encoded into the vectors $\vb a'$. Note that $\mathbf A$ and $\mathbf P$ are diagonal matrices, but $\vb A'$ and $\vb P'$ are not.

Since $\vb P$ was positive definite, $\vb P'$ will also be positive definite~\cite{golub1970stationary}. Therefore, we can diagonalize $\vb P'$ and form the matrix $\vb P'^{-1/2}$. Then, we can write $\vb x = \vb P'^{1/2}\vb a'$, so that the optimization problem reduces to
\begin{align}
    \text{maximize}   &\quad \abs{\frac{\vb x^T\vb P'^{-1/2}\vb A'\vb P'^{-1/2}\vb x}{\vb x^T\vb x}}.
\end{align}
This construction is (the absolute value of) the Rayleigh quotient of the matrix $\vb P'^{-1/2}\vb A'\vb P'^{-1/2}$. The Rayleigh quotient of a symmetric matrix is lower-bounded by its smallest eigenvalue and upper-bounded by its largest eigenvalue, so our optimization problem reduces to the eigenvalue problem for the matrix $\vb P'^{-1/2}\vb A'\vb P'^{-1/2}$.

We found the matrices $\vb A'$ and $\vb P'$ for arbitrary $N$ for the case of one linear constraint (1 LC) and for two linear constraints (2 LC), then calculated the eigenvalues and eigenvectors of $\vb P'^{-1/2}\vb A'\vb P'^{-1/2}$ numerically using \textit{Mathematica}. The matrices for 1 LC and 2 LC are given in Appendix~\ref{app:examples}, while the numerical results are presented in Section IV.

\section{NUMERICAL RESULTS} \label{sec:results}
\subsection[Qualitative Comparison to Traditional MS Gate]{Qualitative Comparison to Traditional MS Gate} \label{sec:results-am-vs-ms}

\begin{figure}
    \centering
    (a)
    \includegraphics[width=0.45\textwidth]{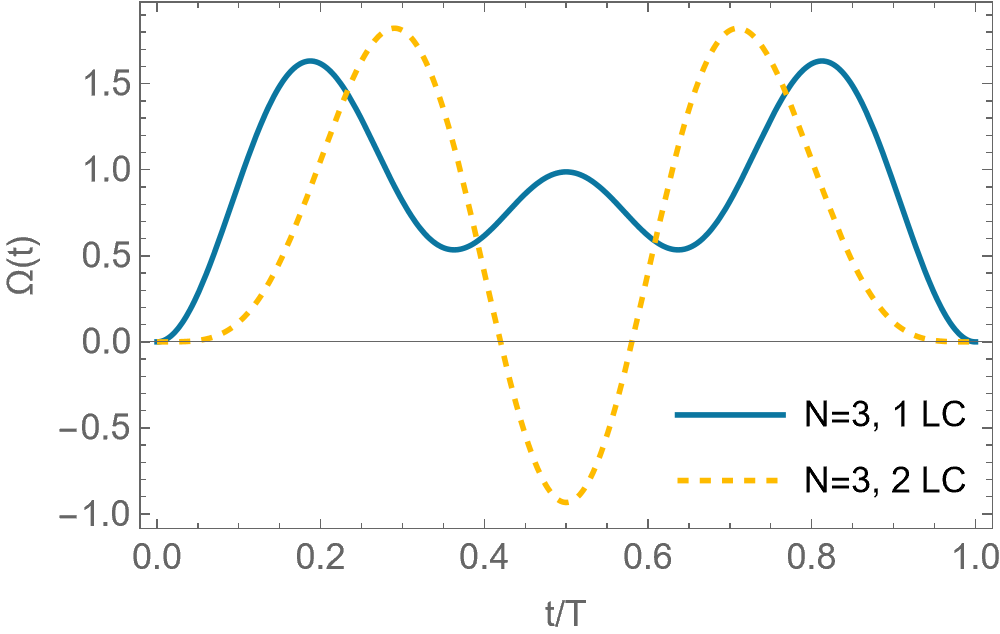}
    (b)
    \includegraphics[width=0.45\textwidth]{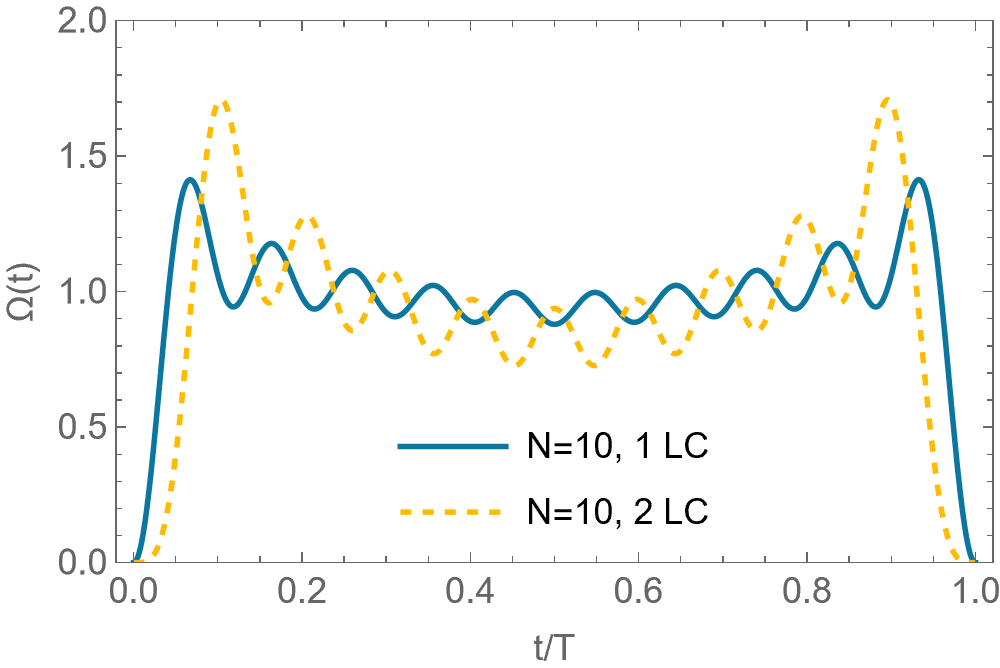}
    (c)
    \includegraphics[width=0.45\textwidth]{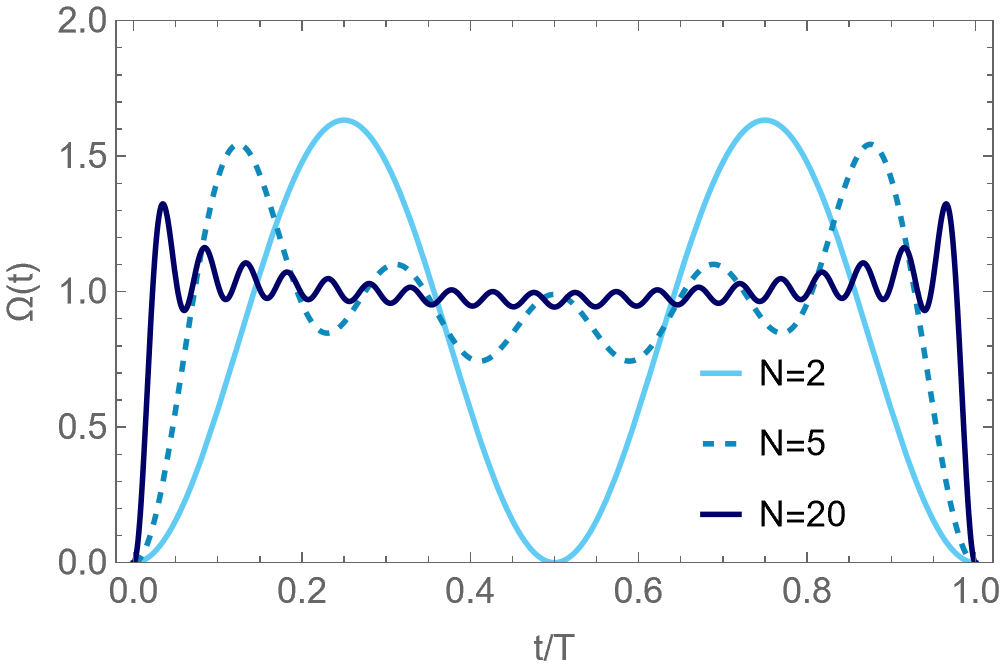}
    \caption{Pulse envelopes of the power-optimized pulses, where three features are observed: slow initial growth, a sharp peak, and oscillations around a constant amplitude for the middle stage. The optimal pulse for 1 LC is compared to the 2 LC pulse for $N=3$ (a) and $N=10$ (b). The trend for increasing $N$ is shown in (c), using the 1 LC case.}
    \label{fig:pulse-envelopes}
\end{figure}

\begin{figure}
    \centering
    (a)
    \includegraphics[width=0.45\textwidth]{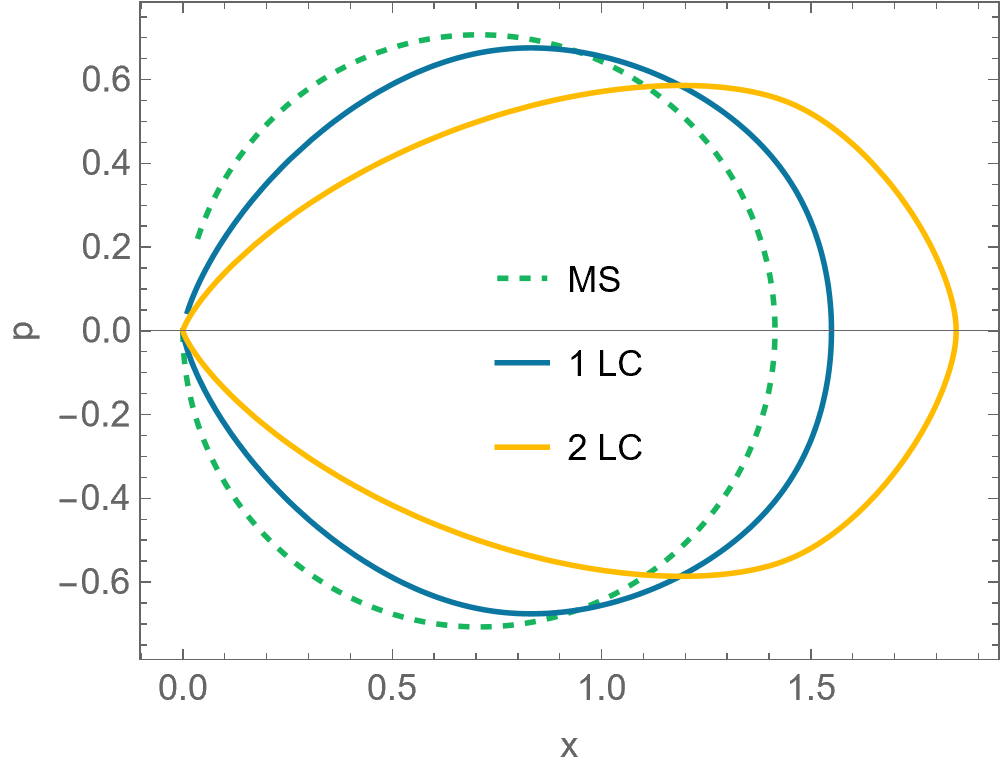}
    (b)
     \includegraphics[width=0.45\textwidth]{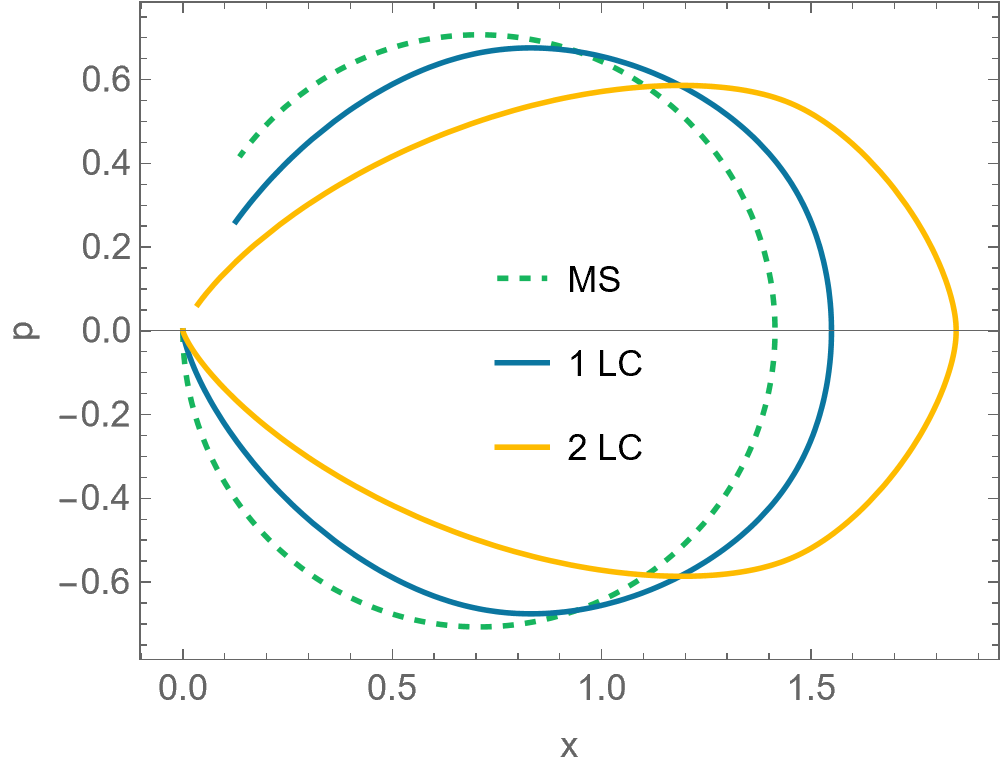}     
    \caption{Trajectories $(G(t),-F(t))$ of MS gate and AM $N=5$ gate with 5\% error (a), and 10\% error (b). The AM gate trajectory is traced out at a non-uniform rate, moving slowly near the origin. This is the source of the robustness to timing errors.}
    \label{fig:error-trajectories}
\end{figure}

Now that the optimal pulse coefficients are calculated, we numerically compare the AM-MS gate and the traditional MS gate. The time evolution of each of these gates is represented in three ways: by the shape of the pulse envelope function $\Omega(t)$, the phase space trajectory, and the population vs. time graph. The plots of $\Omega(t)$, Figure~\ref{fig:pulse-envelopes}, show the pulse envelopes directly and highlight key differences from the constant-amplitude MS gate. The plots of the phase space trajectories (Figures~\ref{fig:trajectories-MS,AM_comparison} and~\ref{fig:error-trajectories}) show the coupling between the electronic states and the motional mode, while the population vs. time graphs, Figures~\ref{fig:gs-pops-MS,AM_comparison} and~\ref{fig:th2-pops-MS,AM_comparison}, show how the probabilities of the $\ket{gg}$ and $\ket{ee}$ states evolve in time.

In Figure~\ref{fig:pulse-envelopes}, we observe that the power-optimized AM pulses all display the significant feature of a soft start followed by a sharp peak. The soft start around $t=0$ (and slow decay toward $t=T$ due to the symmetry of the pulse) corresponds to a slow evolution of the trajectory near the origin in the phase-space picture, improving the sensitivity to timing errors. This feature is the source of the fidelity improvement. In Figure~\ref{fig:pulse-envelopes} (a) and (b),  we observe that the initial growth is slower for additional linear constraints when $N$ is fixed, which agrees with the required robustness to gate-timing errors. As $N$ is independently increased, the initial growth becomes faster, the height of the sharp peak is smaller, and the middle stage of the pulse approximates a constant envelope with smaller variations. The power-optimized AM pulse approaches a constant pulse, a feature of the traditional MS gate, except around the start and the end of the pulse.

\begin{figure}
    \centering
    \includegraphics[width=0.45\textwidth]{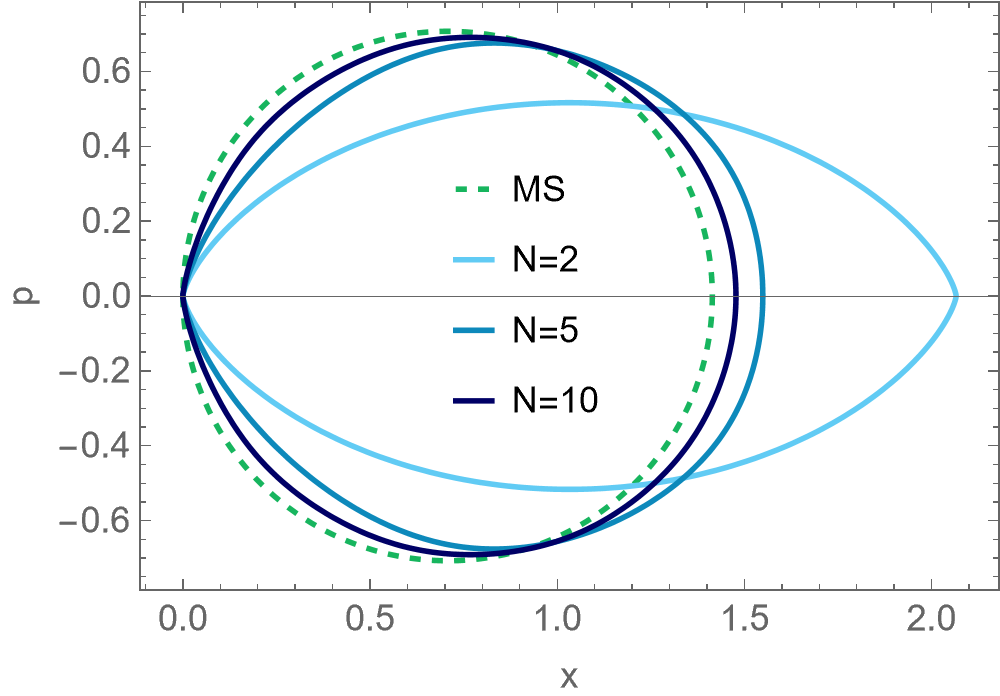}
    \caption{Trajectory $(G(t),-F(t))$ of MS gate compared to AM (with 1 linear constraint) gates for small $N$. The shape of the AM gate approaches the circular trajectory of the standard MS gate as $N$ is increased.}
    \label{fig:trajectories-MS,AM_comparison}
\end{figure}

The phase-space trajectories of the traditional MS gate and the power-optimized AM-MS gate are compared in Figures~\ref{fig:trajectories-MS,AM_comparison} and~\ref{fig:error-trajectories}. As shown in Figure~\ref{fig:trajectories-MS,AM_comparison}, the traditional MS gate traces out a circle intersecting the origin while the power-optimized AM-MS pulses follow elongated trajectories. As the order of $N$ increases in the AM-MS pulses, the trajectory approaches that of the MS gate. The gate timing error $\Delta t$ can be visualized by plotting the trajectory for the range $[0,T-\Delta t]$ in Figure ~\ref{fig:error-trajectories}. The $N=5$ AM gate is compared to the MS gate using 5\%, and 10\% errors for 1 LC and 2 LC. From Figure~\ref{fig:error-trajectories}, we observe that the soft start of the AM-MS gates significantly reduces the residual displacement in phase space for reasonable timing errors, implying a reduction in residual spin-motion entanglement.

\begin{figure}
    \centering
    (a)
    \includegraphics[width=0.45\textwidth]{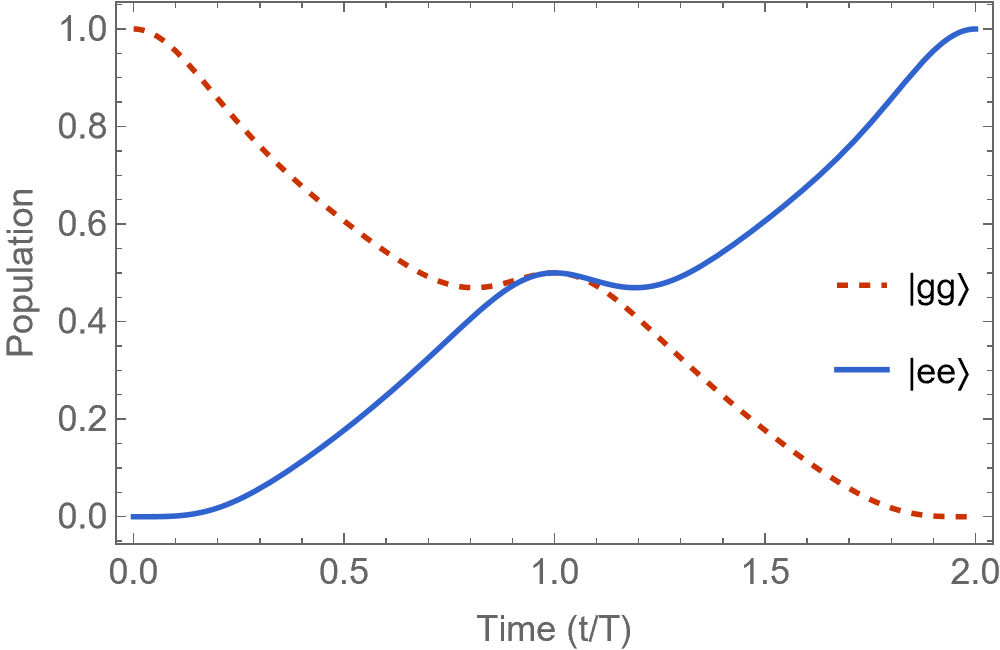}
    (b)
    \includegraphics[width=0.45\textwidth]{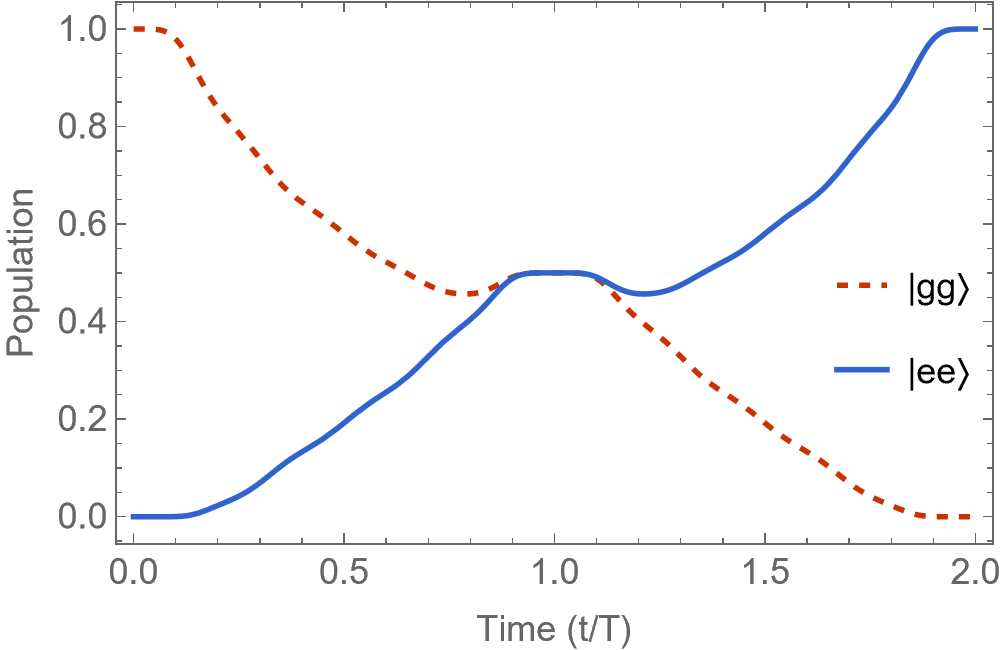}
    \caption{The population dynamics of MS gate (a) compared to AM $N=5$ gate (b) when the motional mode is cooled to the ground state.}
    \label{fig:gs-pops-MS,AM_comparison}
\end{figure}

The population vs. time graph for these two types of pulses is shown in Figure~\ref{fig:gs-pops-MS,AM_comparison}, assuming the system is in the motional ground state, and in Figure~\ref{fig:th2-pops-MS,AM_comparison}, with the system in a thermal state cooled to $\bar{n} = 2$~\footnote{We use the analytic expressions for the population given in \cite{ms-pra} and the probability distribution of a thermal state from \cite{scully_zubairy_1997} to produce these plots.}. Our focus is at one gate time ($t/T = 1$), where we expect an equal coherent superposition of $\ket{gg}$ and $\ket{ee}$.
We use the $N=5$ pulse as a representative for the power-optimized AM-MS gate in these plots. We observe that the amplitude-modulated pulses produce a flatter curve in the region around the gate time compared to the MS gate in both cases, indicating an increased resilience to gate-timing errors. Moreover, this flatness is more evident for thermal states, as the population curve produced by the MS gate becomes sharper around the gate time as the phonon number increases. Note that the population curves do not show the coherence of the state, which is relevant, because tracing out the motional modes in the final state may introduce some decoherence. In the following subsection, we plot the infidelity directly to account for this.

\begin{figure}
    \centering
    (a)
    \includegraphics[width=0.45\textwidth]{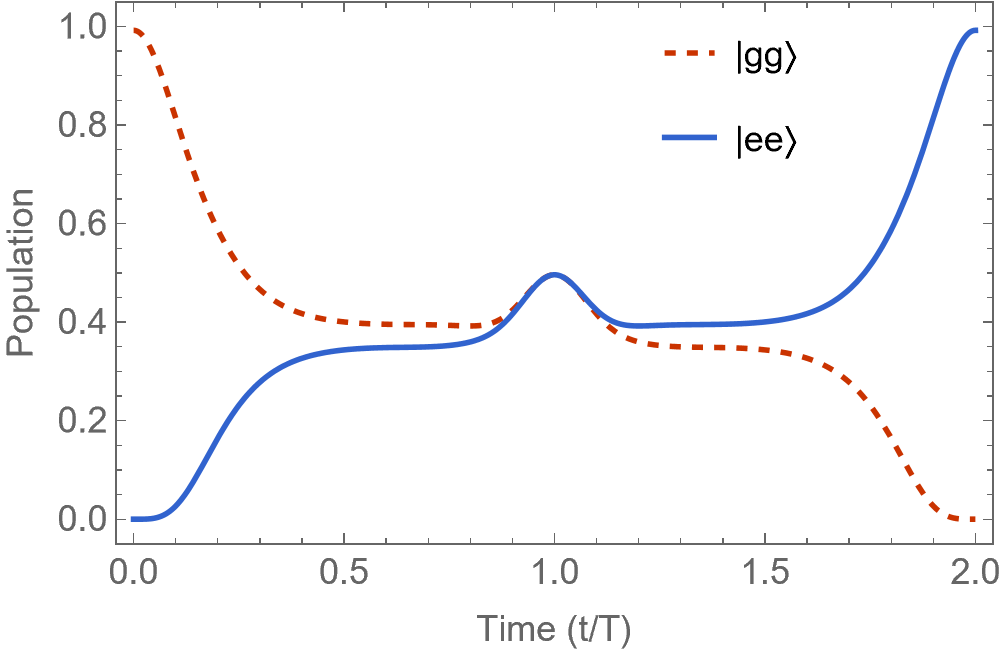}
    (b)
    \includegraphics[width=0.45\textwidth]{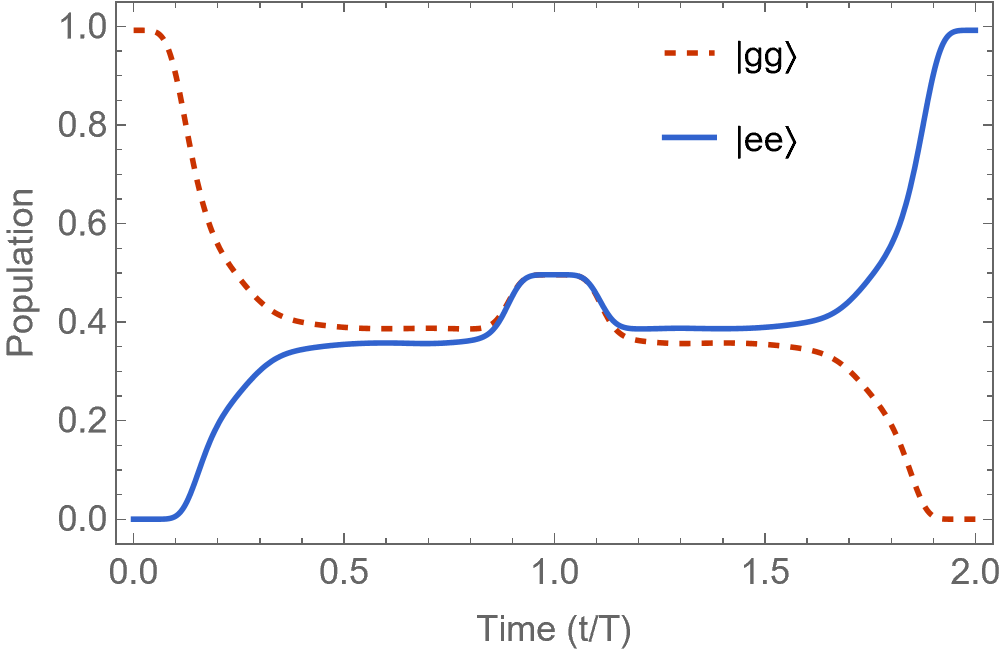}
    \caption{The population dynamics of MS gate (a) compared to AM $N=5$ gate (b) when the motional mode is cooled to a thermal state with the mean phonon number $\bar{n}=2$.}
    \label{fig:th2-pops-MS,AM_comparison}
\end{figure}

\subsection[Gate Fidelity Benefit vs. Laser Power Cost]{Gate Fidelity Benefit vs. Laser Power Cost}
\begin{figure}
    \centering
    (a)
    \includegraphics[width=8cm]{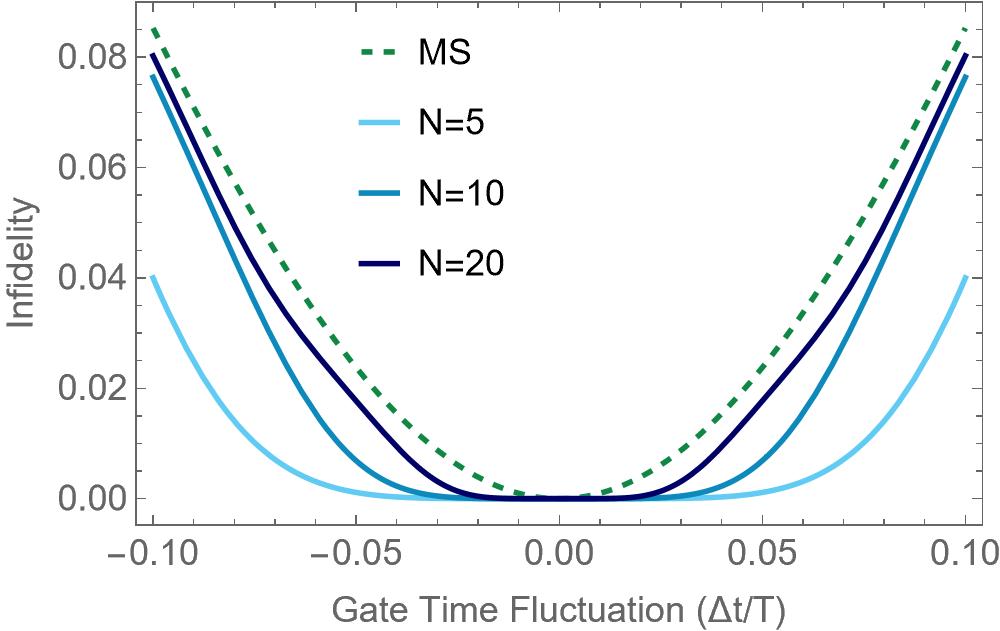}
    (b)
    \includegraphics[width=8cm]{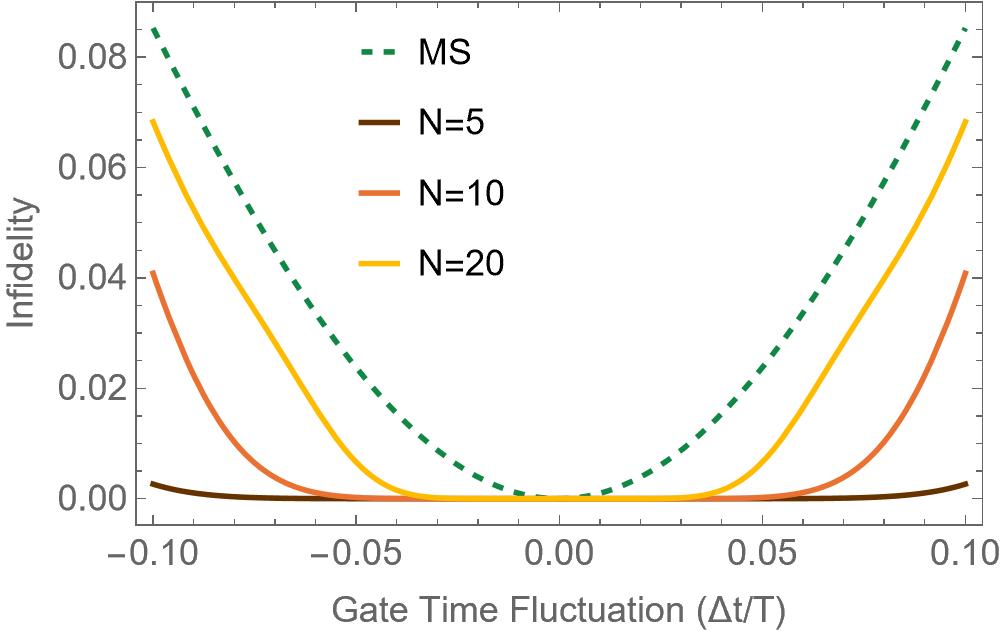}
    \caption{Error rate comparison between constant pulse and Fourier pulses for one linear constraint (a) and two linear constraints (b). Note the flatness around $\Delta t/T=0$ for each AM pulse. The leading contribution of $\Delta t$ to the infidelity is (a) $\order{\Delta t^6}$ and (b) $\order{\Delta t^{10}}$, compared to the leading order $\order{\Delta t^2}$ of the MS pulse.}
    \label{fig:errors-vs-time}
\end{figure}

\begin{figure}[t!]
    \centering
    \includegraphics[width=8cm]{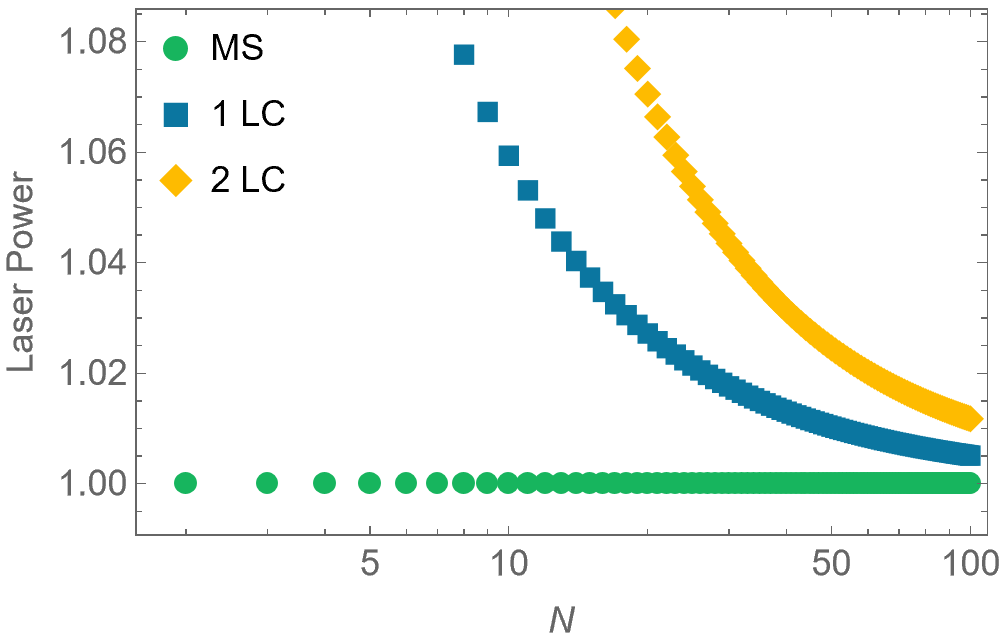}
    \caption{The power required to produce an amplitude modulated pulse with our protocol decreases with increasing $N$. In principle, the complexity of the pulse can be made arbitrarily large~\protect\cite{awg}, resulting in an arbitrarily small increase in the average laser power necessary to implement the entangling gate.}
    \label{fig:power}
\end{figure}

We claimed that our protocol offers fidelity improvements with minimal power cost. Figure~\ref{fig:errors-vs-time} illustrates the former, while figure~\ref{fig:power} offers evidence for the latter.

Figure~\ref{fig:errors-vs-time} directly displays the infidelity $1-\mathcal{F}_{MS}$ with respect to gate time fluctuations ($\Delta t/T$) for 1 LC (a) and 2 LC (b) for a few values of $N$. Both (a) and (b) contain the MS infidelity curve for comparison, which resembles a quadratic function. All AM-MS gates have a ``region of stability" around $\Delta t=0$ where the fidelity is near perfect. For $N=5,10,20$, these regions are approximately $\pm 4\%, \pm 3\%, \pm 1.5\%$, respectively, for the 1 LC case. For the same $N$, they are approximately $\pm 7\%, \pm 5\%, \pm 3\%$, respectively, for the 2 LC case. For a given $N$, we observe that the region of stability is larger with more LCs. However, for a fixed number of LCs, the region of stability narrows as $N$ increases.

Figure~\ref{fig:power} is a plot of the average laser power vs. $N$ for the 1 LC case and the 2 LC case. Note that the MS gate doesn't depend on $N$; it is included as a benchmark. The required power decreases with increasing $N$ for both 1 LC and 2 LC situations. A pulse satisfying 2 LCs costs more power than one satisfying 1 LC, although both appear to asymptotically approach 1. For a pulse with 100 Fourier coefficients, 1 LC can be satisfied with 0.51\% additional power, and 2 LC can be satisfied with 1.2\% additional power. Therefore, using sufficiently large $N$, a more robust pulse can be realized with a minimal increase in laser power from the AM.

\section{CONCLUSION} \label{sec:conclusion}
\subsection{Discussion}
In this article, we explored general amplitude modulation by extending the constant-amplitude laser pulse in the traditional M\o lmer-S\o rensen gate to a Fourier series. This allowed us to derive analytical formulas for the gate trajectory in phase space and gave us a large set of parameters to employ for the power optimization.

Based on the fidelity of the MS gate and its dependence on timing errors, a set of linear constraints on the Fourier coefficients was derived such that the protocol prepares a maximally entangled state with arbitrarily more robustness to gate-timing errors than the original MS protocol. The coefficients were then numerically optimized under the laser power, geometric phase, and first and second linear constraints for varying pulse complexities. Finally, the performance of these pulses was characterized by a qualitative comparison to the MS gate and by a cost-benefit analysis of the relationship between gate fidelity and laser power, which provided evidence that robustness can be gained with minimal extra laser power.

\subsection{Outlook}
Importantly, the protocol developed in this article can be extended to other types of errors. Gate timing errors are a convenient place to begin studying this technique, but these errors are likely not the leading cause of infidelity in the state-of-the-art MS gate~\cite{BallancePRL2016, GTL+16}. By expanding equation~\eqref{fidelity} in other quantities besides gate time (as suggested in~\cite{robust}), the modulated pulses can be constrained by a different system of equations in terms of the Fourier coefficients. Hence, the same strategy can be used on more significant error sources as well, such as laser frequency fluctuations and trap vibrations. In addition, since this protocol achieves a significant asymptotic improvement at vanishing additional average laser power, we expect that it can be used in conjunction with other error-mitigation techniques~\cite{parametric-amplification} to improve two-qubit gates.

Another interesting direction of amplitude modulation is to include the effect of laser amplitude fluctuations~\cite{LO+24}. Not only will these fluctuations affect gate fidelities, but they also fundamentally limit the pulse shaping technique as more precise control of the laser amplitude is required.

Finally, we would also like to relax some of the assumptions in our model to extend the validity of the results to more practical realizations. In particular, neglecting carrier transitions and the third-order Lamb-Dicke terms is understood to contribute coherent errors on the order of $10^{-4}$~\cite{toward}, so these terms will have to be considered in further studies of any entangling gate aiming for a four-nine fidelity.

\section*{Acknowledgements}
We thank Bhargav Das, John Bollinger, and Bryce Bullock for their helpful discussions.

\appendix

\section{Expansion of the Gate Fidelity - Details} \label{app:expansion-details}
The leading order change of the function $\mathcal{F}_{MS}$ is in the first order of $F^2+G^2$. We notice this quantity is nothing but the distance squared of the phase space trajectory to the origin, so let us define
\begin{align*}
    r_p^2 &\equiv F^2+G^2.
\end{align*}
Thus we can expand the exponentials in terms of this quantity, as
\begin{align*}
    e^{-4(\bar n+\half)\frac{r_p^2}{2}} &= 1 - 2\qty(\bar n+\half)\qty(r_p^2) + \order{r_p^2}, \\
    e^{-(\bar n+\half)\frac{r_p^2}{2}} &= 1 - \half\qty(\bar n+\half)\qty(r_p^2) + \order{r_p^2}.
\end{align*}

We can also expand the sine function. Since $A(T+\Delta t)=\pi/2+A(\Delta t)$ as in equation \eqref{AplusDel}, we define $\Delta A = A(\Delta t)$. Thus, we can write
\begin{align*}
    \sin(A+\frac{FG}{2}) &= \cos(\Delta A + \frac{FG}{2}) \\
    &= 1 - \half\qty(\Delta A + \frac{FG}{2})^2 \\ &\tab+ \order{\Delta A + \frac{FG}{2}}^4,
\end{align*}
to leading order. The leading order contribution to gate fidelity is then
\begin{align}
    \mathcal{F}_{MS} &= 1 - \half\qty(\bar n + \half)\qty(F^2+G^2) + \frac{1}{128}\qty(\Delta A + \frac{FG}{2})^2. \label{fidelity-lo}
\end{align}
Note that $A=\order{\Delta t^2}$, so $\left(\Delta A + \frac{FG}{2}\right) = \order{\Delta t^4}$ is small compared to the first term. Hence, the $\qty(\Delta A+\frac{FG}{2})$ term drops out. We proceed by analyzing the leading order change in fidelity,
\begin{align}
    \Delta \mathcal{F}_{MS} &= - \half\qty(\bar n + \half)\qty(F^2+G^2). \label{fidelity-lo-d}
\end{align}
Note that the infidelity---defined as $1-\mathcal{F}_{MS}$---is proportional to the quantity $F^2+G^2$ to the leading order. Therefore, the fidelity is equally sensitive to changes in $F$ and $G$ in general.

\section{Derivatives of the coupling amplitude envelope function $\Omega(t)$} \label{app:Omega-derivatives}
The $k$th derivative is given by
\begin{align*}
    \Omega^{(k)}(t) &= \delta_{k0}\frac{a_0}{2} + \sum_{n=1}^N a_n(n\xi_0)^k\cos(n\xi_0t+k\pi/2) \\&\tab+ \sum_{n=1}^N b_n(n\xi_0)^k\cos(n\xi_0t + (k-1)\pi/2),
\end{align*}
where $\delta_{k0}$ is the Kronecker delta, which we use to compact the notation. When evaluated at $t=T=2\pi/\xi_0$, the sine terms drop out. Thus, we can write the general expression as the piecewise function
\begin{align}
    \Omega^{(k)}(T) &= \begin{dcases}
        \delta_{k0}\frac{a_0}{2} + (-1)^\frac{k}{2}\sum_{n=1}^N a_n(n\xi_0)^k, & \text{even } k, \\
        (-1)^\frac{k-1}{2}\sum_{n=1}^N b_n(n\xi_0)^k, & \text{odd } k.
    \end{dcases} \label{Omega-k-T}
\end{align}
Note that $k$th derivatives of $f$ and $g$ depend on the $(k-l)$th derivatives of $\Omega$. Instead of $\Omega^{(k)}(T)$ itself, we substitute $\Omega^{(k-l)}(T)$ into $f^{(k)}$ and $g^{(k)}$.
Since $F^{(k)}=f^{(k-1)}$ (and likewise for $G$), the substitution is for $\Omega^{(k-1-l)}(T)$. 
However, these $\Omega^{(k-1-l)}(T)$ are each multiplied by $l$th derivatives of either sine or cosine evaluated at the gate time. Thus, only every other term, $\Omega^{(k-1-2l)}(T)$, contributes to each sum. The symbol $l$ is kept as the dummy index even though its meaning has changed. These are expressed as
\begin{widetext}
\begin{align}
    \Omega^{(k-1-2l)}(T) &= \begin{dcases}
        \delta_{0,(k-1-2l)}\frac{a_0}{2} + (-1)^{\frac{k-1}{2}+l}\sum_{n=1}^N a_n(n\xi_0)^{k-1-2l}, & \text{odd } k, \\
        (-1)^{\frac{k-2}{2}+l}\sum_{n=1}^N b_n(n\xi_0)^{k-1-2l}, & \text{even } k>0.
    \end{dcases} \label{Omega-k-1-2l-T}
\end{align}
\end{widetext}

\section{Reduced Linear Constraints} \label{app:LCs}
We show below the derivation of eqs.~(\ref{red-odd-l}, \ref{red-even-l}) from eqs.~(\ref{f-constraint}, \ref{g-constraint}). For $k=1$, we have the following condition
\begin{align}
    F^{(1)}(T) &= 0 & &\implies & \frac{a_0}{2}+\sum_{n=1}^N a_n &= 0, \label{f1}
\end{align}
while $G^{(1)}(T)=0$ for all $a_n,b_n$. For $k=2$, we have
\begin{align}
    F^{(2)}(T) &= 0 & &\implies & \sum_{n=1}^N b_n n &= 0, \label{f2} \\
    G^{(2)}(T) &= 0 & &\implies & \frac{a_0}{2} + \sum_{n=1}^N a_n &= 0. \label{g2}
\end{align}
For $k=3$, we have
\begin{align}
    F^{(3)}(T) &= 0 & &\implies & \frac{a_0}{2} + \sum_{n=1}^N a_n(n^2+1) &= 0, \label{f3} \\
    G^{(3)}(T) &= 0 & &\implies & \sum_{n=1}^N b_n n &= 0. \label{g3}
\end{align}
Finally, we look at $k=4$. We have
\begin{align}
    F^{(4)}(T) &= 0 & &\implies & \sum_{n=1}^N b_n(n^3+3n) &= 0, \label{f4} \\
    G^{(4)}(T) &= 0 & &\implies & \frac{a_0}{2} + \sum_{n=1}^Na_n(3n^2+1) &= 0. \label{g4}
\end{align}
Note that if \eqref{f1} holds, then \eqref{f3} becomes
\begin{align}
    0 &= \underbrace{\frac{a_0}{2} + \sum_{n=1}^N a_n}_\text{=0, by \eqref{f1}} + \sum_{n=1}^N a_n n^2 \nonumber \\
    0 &= \sum_{n=1}^N a_n n^2. \label{red-3}
\end{align}
This is of the form \eqref{red-odd-l}. And now, we show that $G^{(4)}(T)=0$ is not an independent condition from $F^{(1)}(T)=F^{(3)}(T)=0$. From the definition \eqref{G-i}, we write
\begin{align*}
    G^{(4)}(T) &\propto \underbrace{\frac{a_0}{2} + \sum_{n=1}^N a_n}_\text{=0, by \eqref{f1}} + 3\underbrace{\sum_{n=1}^N a_n n^2}_\text{=0, by \eqref{red-3}} \\
    &= 0,
\end{align*}
so we can determine $G^{(4)}(T)=0$ from lower order conditions for $F$.

This trend continues as $k$ increases, meaning that the conditions \eqref{red-odd-l},\eqref{red-even-l} are sufficient to guarantee a pulse with $\order{\Delta t^{2k+2}}$ infidelity contribution. Thus the $G^{(i)}(T) = 0$ conditions can be completely determined from the $F^{(i)}(T) = 0$ constraints. We can arrive at the reduced constraints in Eqs.~(\ref{red-odd-l},\ref{red-even-l}).

\section{Setting $b_i=0$} \label{app:b=0}
The full expressions for average power $P$ and geometric phase $A$ are 
\begin{align}
    P &= \frac{1}{4}a_0^2 + \half\sum_{n=2}^N a_n^2 + b_n^2, \label{power-ab}
\end{align}
and
\begin{align}
    A &\equiv \frac{\xi_0}{T\eta^2} A(T) = -\frac{1}{4}a_0^2 + \half \sum_{n=2}^N \frac{a_n^2+b_n^2}{n^2-1}. \label{phase-ab}
\end{align}

The $b_i$ are taken to be 0 for the remainder of this work. Recall that the MS protocol is the special case where $a_0=2$, meaning that $A = P = 1$. This is the maximal value of the ratio $A / P$ for our construction. Hence, we expect the largest contribution to $A$ to come from the $a_0$ term in the AM-MS protocol as well. Thus, we make an ansatz that the optimal $A$ given a fixed power budget will be negative, noting that only the absolute value of the geometric phase is important in generating a maximally entangled state. Since the $a_i$ and $b_i$ ($i>0$) add positive contributions to $A$, they counteract this larger $a_0$ term. When the first linear constraint [equation \eqref{f1}] is invoked, the fidelity of the operation improves at the cost of laser power investment into the parameters $a_i$ ($i>0$).  

In contrast, nonzero $b_i$ coefficients are not necessary to improve the fidelity. Power investment into these parameters is wasteful because they do not appear in the $k=1$ constraint with $a_0$. In fact, they do not appear in a constraint with $a_0$ for any $k$, so we can likely ignore them for higher $k$ as well. 

Another idea is to set the $a_i=0$ and use only the unconstrained $b_i$ coefficients, so this case is considered. Given some investment of power into the $b_i$ coefficients, the upper bound of the contribution of those coefficients to the geometric phase is 1/3. This is because
\begin{align*}
    P &= \half\sum_{n=2}^N b_n^2, 
\end{align*}
and 
\begin{align*}
    A = \half\sum_{n=2}^N \frac{b_n^2}{n^2-1} &\leq \half\sum_{n=2}^N \frac{b_n^2}{3} = P/3.
\end{align*}
This condition can be used to quantitatively justify the ansatz above. If we obtain a ratio $A/P > 1/3$ with this assumption, it is assured that the chosen coefficients outperform the case with nonzero $b_n$. 

In summary, the $a_0$ term contributes the most to the (absolute value of) geometric phase, while the other $a_n,\ b_n$ coefficients counteract this contribution. The $a_n$ coefficients are indispensable when $a_0$ is nonzero, but the $b_n$ are not constrained by the value of $a_0$ for any $k$. Therefore, we choose $b_n=0$ and choose to optimize only the $a_n$. The $b_n$ coefficients are discussed in this article because expanding the fidelity in terms of errors in parameters other than the gate time will enforce a different system of linear constraints. In other cases, the $b_i$ coefficients may be related to $a_0$ in these constraints, so they must be kept as parameters.

\section[Matrices for 6th and 10th Order Sensitivity]{Example: Matrices for $\order{\Delta \lowercase{t}^6}$ and $\order{\Delta t^{10}}$ Sensitivity} \label{app:examples}
We begin with the case of a fidelity of order $\order{\Delta t^6}$. The pulse is analyzed at an arbitrary value of $N$ to see how the pulse complexity affects the laser power at the same level of robustness.

The first linear constraint gives us a matrix for the laser power which is easy to diagonalize for any $N$. The constraint we have, equation \eqref{f1}, yields
\begin{align}
    a_0 &= -2 \sum_{n=2}^N a_n. \label{a_0,k=1}
\end{align}
Substituting this into equation \eqref{power-a} gives
\begin{align*}
    P &= \left( \sum_{n=2}^N a_n \right)^2 + \half \sum_{n=2}^N a_n^2.
\end{align*}
In the matrix form, the first term corresponds to the matrix of ones $J_{N-1}$ of dimension $N-1$ (i.e. $[J_{N-1}]_{ij}=1$) while the second term is the diagonal matrix $\half \id_{N-1}$, where $\id_{N-1}$ is the identity matrix of dimension $N-1$. Thus, the reduced matrix, $\vb P'$, will have entries $[\vb P']_{ij}=1+(1/2)\delta_{ij}$, with $i,j\in [2,N]\subset\mathbb{Z}$.
By substitution of equation \eqref{a_0,k=1} into equation \eqref{phase-a}, the entries of $\vb A'$ are found to be $[\vb A']_{ij}=-1+(1/(2(j^2-1)))\delta_{ij}$, with $i,j\in [2,N]\subset\mathbb{Z}$.

This was as far as we considered analytically. The rest of the optimization for 1 LC was performed numerically using \textit{Mathematica}. The matrices were constructed and the maximum eigenvalue was found. The general forms of $F$ \eqref{F-major} and $G$ \eqref{G-major} were used to speed up the calculations required in the plots. The results of the numerical work are presented in Section IV.

For $\order{\Delta t^{10}}$, we followed a similar strategy. The linear system was solved, and the reduced coefficients $a_0,a_2$ were substituted into equations \eqref{power-a} and \eqref{phase-a}. From these expressions, the matrices $\vb A'$ and $\vb P'$ were found. The constraints used are
\begin{align*}
    \frac{a_0}{2} + \sum_{n=2}^N a_n &= 0, \\
    \sum_{n=2}^N a_n n^2 &= 0.
\end{align*}
The first equation again gives \eqref{a_0,k=1}, while the new constraint allows us to write
\begin{align}
    a_2 &= -\frac{1}{4} \sum_{n=3}^N a_n n^2. \label{a2-k34}
\end{align}
Thus,
\begin{align}
    a_0 &= -2a_2 - 2\sum_{n=3}^N a_n \nonumber \\
    &= -2 \left(-\frac{1}{4} \sum_{n=3}^N a_n n^2\right) - 2\sum_{n=3}^N a_n \nonumber \\
    &= \half \sum_{n=3}^N a_n (n^2-4). \label{a0-k34}
\end{align}
Now, substituting both~\eqref{a0-k34} and~\eqref{a2-k34} into~\eqref{power-a} yields
\begin{align*}
    P &= \frac{1}{16}\left(\sum_{n=3}^N a_n(n^2-4)\right)^2 \\&\tab+ \frac{1}{32}\left(\sum_{n=3}^N a_n n^2\right)^2 + \half \sum_{n=3}^N a_n^2,
\end{align*}
which results in a reduced matrix $\vb P'$ with entries
\begin{align}
    [\vb P']_{ij} &= \frac{1}{16} (i^2-4)(j^2-4) + \frac{1}{32} i^2j^2 + \half \delta_{ij}, \label{B-prime-k34}
\end{align}
with $i,j$ indexed from 3 to $N$. Similarly, substituting~\eqref{a0-k34} and~\eqref{a2-k34} into~\eqref{phase-a} yields
\begin{align*}
    A &= -\frac{1}{16}\left(\sum_{n=3}^N a_n(n^2-4)\right)^2 \\&\tab+ \frac{1}{96}\left(\sum_{n=3}^N a_n n^2\right)^2 + \half\sum_{n=3}^N \frac{a_n^2}{n^2-1},
\end{align*}
which gives a reduced matrix $\vb A'$ with entries
\begin{align}
    [\vb A']_{ij} &= -\frac{1}{16}(i^2-4)(j^2-4) + \frac{1}{96}i^2j^2 + \frac{1}{2(j^2-1)}\delta_{ij}, \label{A-prime-k34}
\end{align}
with $i,j$ again indexed from 3 to $N$. These matrices were again constructed in \textit{Mathematica}, and the maximum eigenvalue of $\vb P'^{-1/2}\vb A'\vb P'^{-1/2}$ was found along with its corresponding eigenvector. The pulse coefficients were recovered from this eigenvector, allowing the power-optimized pulses to be studied in detail. The results are portrayed in Section IV.

\end{document}